\documentclass[12pt]{article}
\begin{document}

\author{Pu Xu$^1$,Jia-Lun Ping$^2$, 
        Fan Wang$^3$, T.Goldman$^4$}

\title{QED-QCD Interference Effect on the Charge Dependent N-N Interaction}

\setcounter{page}{0}

\maketitle

\vspace*{2.5in}
\begin{abstract}
The charge symmetry breaking and charge independence breaking N-N
$^1$S$_0$ scattering length differences, $\Delta a_{CSB}$ and $\Delta
a_{CIB}$, are calculated by a resonating group method with a quark
cluster model. By adding the QED-QCD interference effect to the quark
mass difference and the electromagnetic interaction, the $\Delta
a_{CSB}$ and $\Delta a_{CIB}$ values can be reproduced with model
parameters constrained by the hadron isomultiplet mass splitting.
\end{abstract}

\vspace*{-4.1in}
\begin{center}
$^1$ Physics Department, Nanjing University of Science 
and Technology, Nanjing, 210094,China\\
$^2$Physics Department, Nanjing Normal University, Nanjing 210097,China\\
$^3$ Physics Department and Center for Theoretical physics, Nanjing
University, Nanjing 210093,China\\
$^4$Theoretical Division, Los Alamos National Laboratory, 
Los Alamos, NM 87545, USA
\end{center}

\vspace*{-4.1in}
\begin{flushright}
LA-UR-98-1336\\
hep-ph/9911008\\
\end{flushright}
\vspace*{7.0in}


\pagebreak

There are two main approaches to hadronic interactions: One is the
meson exchange model, the other, the quark cluster model. The meson
exchange model\cite{1} fits the hadron interactions quantitatively with
the aid of many phenomenological meson-baryon coupling constants and
form factors (or short range phenomenologies). Quark cluster
models\cite{2} bring a deeper understanding of the short range
repulsion but most of them can not produce the N-N intermediate range
attraction and must invoke effective meson exchange again. Wang {\it et
al.}\cite{3} have developed a model which takes into account quark
delocalization and color screening, and so is able to reproduce the N-N
short range repulsion and the intermediate range attraction
simultaneously. It also draws an analogy between nuclear and molecular
forces.

For the charge dependent N-N interaction, the meson exchange model
attributes the charge independence breaking mainly to the $\pi$ mass
difference and the charge symmetry breaking mainly to $\rho-\omega$
mixing\cite{4}.  While accepting that the primary origins of the charge
dependence of the hadronic interaction are the quark mass difference
and the electromagnetic interaction, meson exchange model practitioners
believe that these two effects play their role indirectly: The
electromagnetic interaction affects the $\pi$ masses and the quark mass
difference induces the $\rho-\omega$ mixing. However, as pointed out by
S.N. Yang and Pauchy W.-Y. Hwang\cite{5}, such an optimistic picture
for the meson exchange model fit of the $\Delta a_{CSB}$ and $\Delta
a_{CIB}$ is questionable. Different calculations of the
$\pi^{\pm}-\pi^{0}$ mass difference effect on $\Delta a_{CIB}$ give
quite different results ranging from $2.64\pm 0.16~fm$ \cite{6},
through $3.11\pm 0.1~fm$\cite{7} to $3.6\pm 0.15~fm$\cite{8}. The
$\gamma\pi$ contribution estimates are even more diverse, ranging from
$-0.53~fm$\cite{9} to $0.67~fm$\cite{10} and a value of $1.1~fm$ is
listed in table 3.3 of reference\cite{4}. This is an average of two
calculations with even more significantly different $\Delta
a_{CIB}\left(\gamma \pi\right)$ values. Moreover, it has been argued
that the $\rho-\omega$ mixing is strongly momentum transfer dependent
and therefore the charge symmetry breaking in $N-N$ scattering due to
$\rho-\omega$ mixing is strongly suppressed\cite{11}.

There are also direct quark model approaches to the charge dependent
N-N interaction\cite{12,13}, wherein the quark mass difference effect
and the electromagnetic interaction have been taken into account. It
seems generically impossible to fit the charge symmetry and charge
independence breaking simultaneously in this kind of approach, because
the quark mass difference does not affect the $n$-$p$ scattering and
the electromagnetic effect is not large enough\cite{12} to reproduce
the charge independence breaking.  A related concern is that the quark
mass difference is not directly observable, because the quarks are
permanently confined. They are principally determined using hadron
isomultiplet masses. A quark mass difference so determined is in turn
dependent on the assumption of the primary origin of the charge
dependence. Goldman, Maltman and Stephenson (GMS)\cite{14} proposed
that "QED-QCD interference" induces a charge dependent q-q interaction
effect not taken into account in previous considerations.  If this
effect is included in the hadron isomultiplet mass fitting, GMS can fit
the existing hadron isomultiplet masses (with large experimental
errors) with a quark mass difference $-$4.96 MeV $\leq\Delta
m(=m_d-m_u) \leq$ 6.28 MeV. Their most favorable parameter set has
$\Delta m$=0.66 MeV, which is quite different from the $\Delta m \sim$
3MeV recommended in Ref.~\cite{4} and the value used is quark model
calculations\cite{12,13}. If the QED-QCD interference effect is really
as important as proposed by GMS, this interference effect should be
added to the quark mass difference and electromagnetic interaction and
the charge symmetry breaking in nuclear and particle physics
reanalyzed.  The QED-QCD interference effect contributes to both charge
symmetry and charge independence breaking. Our interest here is to
check whether one can produce the charge symmetry and charge
independence breaking appearing in the N-N interaction simultaneously
in a quark model approach by taking the quark mass difference, the
electromagnetic interaction and the QED-QCD interference into account
together.

It has been proven that for a color singlet system without explicit
gluon excitations, the quark degree of freedom description and the
colorless meson baryon degree of freedom descriptions are
equivalent\cite{15}. For a real calculation, however, the Hilbert space
is always truncated. One then faces the question of which
truncation--quark approach or meson baryon approach--is more
efficient.  Similarly, it is also worth considering which one--quark or
meson baryon approach--is more efficient for the description of the
charge dependent N-N interaction. It is well known that $\rho-\omega$
meson exchange is responsible for both the short range repulsion and
the spin-orbit force of the N-N interaction. On the other hand, it has
been shown that the N-N short range repulsion\cite{2} and the
spin-orbit splitting\cite{16} can be deduced in the quark model
approach.  Therefore one expects that it should be possible to replace
the $\rho-\omega$ mixing by a direct quark model description. Since the
$N-N$ intermediate range attraction described by pion exchange can also
be described by quark delocalization and color screening, it is
interesting to check whether the pion mass difference effect can be
described by a quark effect as well.

Motivated by the reasoning above, we have carried out a quark
cluster model calculation of the N-N $^1$S$_0$ scattering length
differences $\Delta a_{CSB}=a_{pp}-a_{nn}$ and $\Delta a_{CIB}=\frac
12(a_{pp}+a_{nn})-a_{np}$, where $a_{pp (nn,np)}$ is the $pp (nn,np)$
$^1$S$_0$ scattering length. The u,d quark mass difference $\Delta
m=m_{d}-m_{u}$, the q-q electromagnetic interaction $V_{qq}^{\gamma}$
and the QCD-QED interference induced q-q interaction $V_{qq}^I$ are all
taken into account in this calculation.

A resonating group method based on the quark cluster model is used to
carry out the N-N scattering length calculation\cite{12}.  The
Hamiltonian of the N-N system is assumed to be
\begin{eqnarray}
& & H(1\cdots 6)=\sum (m_i+\frac{p_i^2}{2m_i})-T_c+\sum
(V_{ij}^C+V_{ij}^G+V_{ij}^\gamma +V_{ij}^I) , \nonumber \\
& & =\sum (m+\frac{p_i^2}{2m})-T_c+\sum (V_{ij}^C+\bar{V}
_{ij}^G)+\sum (\delta m_i+\delta K_i)+\sum (\delta V_{ij}^G+V_{ij}^\gamma
+V_{ij}^I) ,
\end{eqnarray}
where $m_{i}$ is the quark mass, $T_c$ is the total center of mass
kinetic energy, the $V_{ij}$ with superscripts $C, G, \gamma $ and $I$
represent the confinement, gluon exchange, photon exchange and QED-QCD
interference q-q interactions respectively; m is the average of the
u and d quark masses.
\begin{eqnarray}
& & m=\frac{m_u+m_d}2~~~\Delta m=m_{d} - m_u, \nonumber \\
& & \delta m_i=m_i-m=-\frac{\Delta m}2\tau _{iz} 
\end{eqnarray}

\begin{equation}
\label{e3}\delta K_i=(\frac 1{2m}_i-\frac 1{2m})p_i^2=\frac{p_i^2}{2m}
  \frac{\Delta m}{2m}\tau _{iz} 
\end{equation}
\begin{equation}
\label{e4}V_{ij}^C=-a_C \vec{\lambda}_i \cdot \vec{\lambda}_j r_{ij}^2 
\end{equation}
\begin{equation}
\label{e5}V_{ij}^G+V_{ij}^\gamma =(\frac{\alpha_s}4\vec{\lambda}_i
\cdot \vec{\lambda}_j+\alpha Q_iQ_j)(\frac 1{r_{ij}}-\frac
\pi 2(\frac 1{m_i^2}+\frac 1{m_j^2}+\frac{16\vec{S}_i\cdot
\vec{S}_j}{3m_im_j})\delta (\vec{r}_{ij})+\cdots ) 
\end{equation}
$$
\bar{V}_{ij}^G=\frac{\alpha _s}4\vec{\lambda}_i \cdot 
\vec{\lambda}_j (\frac 1{r_{ij}}-\frac \pi {m^2}(1+\frac{8
\vec{S}_i \cdot \vec{S}_j}3)\delta (\vec{r}_{ij})+\cdots ) 
$$
\begin{equation}
\label{e6}\delta V_{ij}^G=V_{ij}^G-\bar{V}_{ij}^G=-\frac{\alpha _s}4
\vec{\lambda}_i\cdot \vec{\lambda}_j\frac \pi
{m^2}(1+\frac{8\vec{S}_i \cdot \vec{S}_j}3)\delta (
\vec{r}_{ij})\frac{\Delta m}{2m}(\tau _{iz}+\tau _{jz}) 
\end{equation}

\begin{eqnarray}
V_{ij}^I & = & \frac 14 
\vec{\lambda}_i\cdot \vec{\lambda}_j
(AQ_iQ_jf_{el}(r_{ij})+BQ_iQ_j\vec{S}_i\cdot \vec{S}_j
f_{mag}(r_{ij})+C(Q_i^2+Q_j^2)f_{el}(r_{ij}) \nonumber \\
& &  +D(Q_i^2+Q_j^2) \vec{S}_i\cdot \vec{S}_j f_{mag}(r_{ij})) 
\end{eqnarray}

$$
\label{e7}f_{el}(r)=\frac{\frac 1r}{<\frac 1r>_{nucleon}},~~~~
f_{mag}(r) =\frac{\frac{8\pi }{3m^2}\delta (\vec{r})}
   {<\frac{8\pi }{3m^2}\delta (\vec{r})>_{nucleon}} 
$$
Here $\alpha_s (\alpha)$ is the quark-gluon (electromagnetic) coupling
constant, $a_C$ is the confinement strength, $Q_i, \vec{S}_i,
\tau_{iz}$ and $\vec{\lambda}_i$ are the quark charge, spin, third
component of isospin and color SU$_3$ generators, respectively; A,B,C
and D are the parameters\cite{14} of the QED-QCD interference induced
quark interactions.

In principle, we can directly use the Hamiltonian (1) and the
resonating group method to get both the charge independent and charge
dependent N-N interactions\cite{13}. In practice, quark model
approaches have not obtained a charge independent N-N interaction as
good as the phenomenological or the meson exchange model
ones\cite{2,3}. To focus this study on the charge dependent N-N
interaction, we assume the charge independent part can be well
described by the Reid soft core phenomenological potential\cite{17}.
The resonating group equation for the N-N scattering is thereby reduced
to\cite{12}
\begin{eqnarray}
& & [\nabla ^2+k^2-\frac{2\mu }{\hbar ^2}(V_{NN}+V_{em})]F({\bf r)}=
\frac{2\mu }{
\hbar ^2}\int K({\bf r,r}^{\prime })F({\bf r}^{\prime }{\bf )}d{\bf r}
^{\prime }  \nonumber \\
& & K({\bf r,r}^{\prime })=-9\langle [\psi _{N_1}(1,2,3)
\psi_{N_2}(4,5,6)]_{ST}\mid \sum_{\alpha} O_\alpha \mid P_{34}[\psi
_{N_1}(1,2,3)\psi _{N_2}(4,5,6)]_{ST}\rangle 
\end{eqnarray}
where the $V_{NN}$ is the Reid soft core potential, $V_{em}$ is the
direct electromagnetic interaction between nucleons, $\mu $ is the
reduced mass $\mu $ =$\frac{M_n+M_p}{4}, \psi _N$ is the quark model
nucleon wave function (color singlet, SU$_4^{\sigma \tau }$ symmetric,
Gaussian orbital with a size parameter b), $P_{34}$ is a quark exchange
operator, $[~]_{ST}$ means the nucleon spin-isospin is coupled to the
channel spin S and isospin T.  The $O_\alpha $ are the five charge
dependent terms: $\delta m_i, \delta K_i, \delta V_{ij}^G,
V_{ij}^\gamma $ and $V_{ij}^I$. The first four terms have been studied
in \cite{12}. Here we concentrate on the effects of the last term, the
QED-QCD interference, V$_{ij}^I$.

We take two approaches to study the effect of $V_{ij}^I$. One is to
follow GMS\cite{14}, i.e., assume that the QED-QCD interference induces
a q-q interaction as shown in Eq.(7), and fix all model parameters (the
average u, d quark mass m, the u, d quark mass difference $\Delta m$,
quark gluon coupling constant $\alpha _s$, nucleon size b and
parameters A, B, C, D) from an overall fit to baryon isomultiplet mass
splittings. We call this a phenomenological model (MP). Due to the
relatively large errors involved in the experimental hadron masses used
in the overall fitting, especially those of $\Sigma^{*}$ and
$\Sigma_{c},$ GMS tested 135 sets of parameters. The results listed in
Table I, labeled by MP, correspond to their most favorable set. Other
parameter sets have been tested as well, but not $\Delta m<0$ cases,
because those are unacceptable from the general point of view on the
quark mass difference~\cite{4}. In Table I, two additional results,
labeled as MP I and MP II, are listed to indicate the level of
sensitivity of the N-N scattering lengths to the variation of the
phenomenological parameters. The other approach is to take the A, B, C
and D parameters from a perturbative QED-QCD calculation~\cite{18},
which includes the vertex electromagnetic penguin, photon and gluon box
and crossed box diagrams. The interference interaction calculated this
way is
$$
V_{ij}^I=\frac 14\vec{\lambda}_i\cdot \vec{\lambda}_j
\alpha _s\alpha \{\frac{16(1-\ln 2)}{4\pi }Q_iQ_j\frac 1{r_{ij}}-\frac{
12(1+\ln 2)}{m^2}Q_iQ_j\vec{S}_i\cdot \vec{S}_j\delta (
\vec{r}_{ij})- 
$$
\begin{equation}
\label{e9}\frac 43(Q_i^2+Q_j^2)\frac{\vec{S}_i\cdot 
\vec{S}_j}{m^2}\delta (\vec{r}_{ij})+\cdots \} 
\end{equation}
There are many other terms in the perturbative QED-QCD interference
calculation result beyond those explicitly shown here. Only the three
terms shown in Eq.(9), which have the same form as assumed by
GMS\cite{14}, have been included in our scattering calculation. As the
QED-QCD interference effect is nonperturbative, we should perform a
nonperturbative calculation.  However from the phenomenological success
of the effective one gluon exchange Breit-Fermi interaction, we expect
that a perturbative calculation of the QED-QCD interference interaction
is useful in the study of the GMS effect, because we are doing a QED
correction to the effective one gluon exchange. We call this a QCD
model (MQ). The parameters other than those that appear in Eq.(9) are
determined by re-fitting the n-p and $\Delta$ mass differences with two
choices of the nucleon size $b$ but fixed $\alpha _s$. They are labeled
as MQ I and MQ II in Table I.


Altogether there are five model results listed in Table I. The
corresponding five parameter sets are listed together in Table II. In
order to show the effect of the individual QED-QCD interference terms,
in the first to fourth rows of Table I, we list the scattering length
corrections, $\Delta a_{ij}=a_{ij}-a_{NN}$, due to each of the A, B, C
and D terms in Eq.(7). The charge independent Reid soft core potential
$V_{NN}$ in Eq.(8) is always included; it gives a charge independent
$a_{NN} = -$17.13 fm. The different terms increase or decrease the
scattering lengths differently.  The fifth row is the correction due to
the coherent sum (A+B+C+D) of the four terms in Eq.(7). The sixth row
lists the scattering length corrections due to the coherent sum of
all other relevant effects: the quark mass difference, gluon and photon
exchange interactions, i.e., $\delta m+\delta K+\delta V^G+V^\gamma .$
The last row, labelled Full, lists the overall results with the five
charge dependent terms of Eq.(1) all included.

It should be noted that the total scattering length correction is not
always equal to the sum of the individual ones, even though the
individual corrections to the scattering length are small.

The main feature we find is that the corrections due to different
charge asymmetry interactions tend to cancel each other and affect the
scattering length coherently.

Because we based our calculation on the Reid soft core which gives a
charge independent a$_{NN} =-$17.13 fm, quite close to the nuclear p-p
scattering length, the only meaningful quantities from our analysis are
the charge symmetry breaking scattering length difference $\Delta
a_{CSB}=a_{pp}-a_{nn}$ and the charge independence breaking scattering
length difference $\Delta a_{CIB}=\frac{1}{2}(a_{pp}+a_{nn})-a_{np}$
(expressed as $\Delta a_{CD}$ in~\cite{4}). Our model results $\Delta
a_{CSB}$ and $\Delta a_{CIB}$ are listed in Table II together with the
corresponding model parameters.

For the GMS phenomenological models, the calculated $\Delta a_{CSB}
(\sim$2-3 fm) is too large in comparison with the value of 1.5 fm
recommended in Ref.~\cite{4}. The $\Delta a_{CIB}$ ($\sim$1 fm) is much
too small in comparison with 5.7 fm recommended in Ref.~\cite{4} and
for the GMS most favorable parameter set it is $-$1.5 fm, which is the
wrong sign. These results are consistent with those of~\cite{19}.

Interestingly, and perhaps even surprisingly in view of their
perturbative origins, the QCD models produce better results. The MQ II
leads to $\Delta a_{CSB}$=2.16 fm, $\Delta a_{CIB}$=7.33 fm, which are
not too different from the recommended values of 1.5 and 5.7 fm
respectively.  As a further demonstration of the plausibility of the
QCD models, in Tables I and II we added an additional result, labelled
as MQP, where the parameters are constrained by the n-p and
$\Delta^{++}-\Delta^{0}$ mass differences.  It gives $\Delta
a_{CSB}$=1.50 fm and $\Delta a_{CIB}$=5.73 fm, almost identical to the
values recommended in~\cite{4}.

We conclude that, taking into account the quark mass difference, the
electromagnetic interaction and the QED-QCD interference effects
together, it is possible, using a quark cluster model, to deduce the
charge dependent effects found in the N-N interaction.  However, with
the present model parameters, the GMS phenomenological model can not
explain the charge asymmetry appearing in the N-N $^1S_0$ scattering
lengths.  The perturbative QCD model appears plausible, but further
quantitative work is needed.

F. Wang and J.L. Ping acknowledge the hospitality of the Theoretical
Division of LANL. This work is supported by the NSF(19675018), SEDC and
SSTC of China and by the US Department of Energy under contract
W-7405-ENG-36.

{\it Note added.} After this work was completed, we learned of
Ref.\cite{vgrb}, which independently finds that heavy baryon mass
splittings cannot be consistently described without the inclusion of
the interference terms studied here (which they refer to as
"electromagnetic penguins"), despite consideration of several other
different possible additional contributions besides the conventional
ones of quark mass and photon exchange effects. The need for
"electromagnetic penguins" in heavy-light systems has also been argued
in Ref.\cite{kissl}.

\newpage

\begin{center}
Table 1. N-N $^{1}S_{0}$ scattering length corrections.

\begin{tabular}{|c|c|c|c|c|c|c|c|} \hline
   &  & MP  & MPI   & MPII  & MQI & MQII  & MQP \\ \hline
   & $\Delta a_{pp}$ & 0.05 & 0.04 & 0.04 & $-$0.03 & $-$0.09 & $-$0.09 \\ 
 A & $\Delta a_{nn}$ & 0.01 & 0.01 & 0.01 & $-$0.00 & $-$0.01 & $-$0.00 \\
   & $\Delta a_{np}$ & 1.58 & 1.24 & 1.08 & $-$1.32 & $-$2.65 & $-$3.82 \\ \hline
   & $\Delta a_{pp}$ & $-$2.45 & $-$2.28 & $-$1.96 & 1.54 & 1.60 & 0.67 \\
 B & $\Delta a_{nn}$ & $-$1.12 & $-$1.05 & $-$0.91 & 0.79 & 0.79 & 0.34 \\
   & $\Delta a_{np}$ & $-$0.38 & $-$0.36 & $-$0.31 & 0.29 & 0.31 & 0.13 \\ \hline
   & $\Delta a_{pp}$ & $-$1.88 & $-$0.5 & 0.30 &         &      & $-$1.01 \\
 C & $\Delta a_{nn}$ & $-$1.53 & $-$0.42 & 0.25 &       &      & $-$0.83\\
   & $\Delta a_{np}$ & $-$1.70 & $-$0.46 & 0.28 &       &      & $-$0.92 \\ \hline
   & $\Delta a_{pp}$ & 2.90 & $-$0.16 & $-$3.18 & 0.17 & 0.20 & 0.98 \\
 D & $\Delta a_{nn}$ & 1.07 & $-$0.05 & $-$0.93 & 0.06 & 0.08 & 0.33 \\
   & $\Delta a_{np}$ & 2.04 & $-$0.11 & $-$1.99 & 0.11 & 0.14 & 0.66 \\   \hline
   & $\Delta a_{pp}$ & $-$0.33 & $-$3.11 & $-$5.40 & 1.65  & 1.69 & 0.67 \\
 A+B+C+D & $\Delta a_{nn}$ & $-$1.42 & $-$1.58 & $-$1.64 & 0.84 & 0.85 & $-$0.12\\
   & $\Delta a_{np}$ & 1.88 & 0.41 & $-$0.70 & $-$0.87 & $-$2.07 & $-$3.91 \\ 
   \hline
   & $\Delta a_{pp}$ & 1.17 & 1.90 & 2.37 & 2.04 & 1.81 & 1.17 \\
 $\delta m+\delta K+\delta V^G+V^{\gamma} $
   & $\Delta a_{nn}$ & 0.20 & $-$0.71 & $-$1.39 & $-$0.90 & $-$0.19 & 0.2\\
   & $\Delta a_{np}$ &$-$0.66 & $-$0.66 & $-$0.66 & $-$0.66 & $-$0.81 & $-$0.66 \\ \hline
   & $\Delta a_{pp}$ & 0.87 & 0.99 & 1.51 & 3.20 & 2.73 & 1.75 \\
 Full & $\Delta a_{nn}$ & $-$1.19 & $-$1.22 & $-$1.94 & $-$0.16 & 0.56 & 0.08\\
   & $\Delta a_{np}$ & 1.36 & $-$1.23 & $-$1.23 & $-$2.64 & $-$5.69 & $-$4.89 \\ 
   \hline
\end{tabular}


\vspace{1cm}

Table 2. Calculated $\Delta a_{CSB}, a_{CIB} $ and the model parameters.

\begin{tabular}{|c|c|c|c|c|c|c|c|c|c|c|}\hline
   & m(MeV) & $\Delta$ m(MeV) & $\alpha_{s}$ & b(fm) & A(MeV) & B(MeV) &
    C(MeV) & D(MeV) & $\Delta a_{CSB}$ & $\Delta a_{CIB}$ \\ \hline
exp. &     &              &            &       &       &       &       &
           & $\sim 1.5 $ & $ \sim 5.7$ \\ \hline
MP & 330   & 0.6609       & 1.624      & 0.617 &$-$1.666 & 5.894 & 5.340 &
    $-$6.258 & 2.06 & $-1.52$ \\ \hline
MPI&330    &3.7724        & 1.624      & 0.617 &$-$1.278 & 5.543 & 1.551 & 
     0.296 & 2.21 & 1.12 \\ \hline
MPII&330   &5.8927       & 1.624      & 0.617 &$-$1.104 &4.844  & $-$0.953 &
     4.861 & 3.45 & 1.02 \\ \hline
MQI &330   &4.37       &1.624      &0.617   &1.182 & $-$4.592 & 0    &
     $-$0.301 & 3.36 & 4.26 \\ \hline
MQII & 330 & 1.91        &1.624       &0.80    &0.911   & $-$2.032 & 0  & 
     $-$0.133 & 2.16 & 7.33 \\ \hline
MQP & 330 & 0.6609       &1.624       &0.617    &2.804   & $-$2.444 & 2.824  & 
     $-$1.040 & 1.50 & 5.73 \\ \hline
\end{tabular}
\end{center}

\end{document}